\documentclass[twocolumn,superscriptaddress,showpacs,10pt,prl]{revtex4}

\usepackage{graphics}%

\begin{document}

\title{Hopping conductivity in the quantum Hall effect -- revival of universal scaling}

\author{F.~Hohls}
\email{hohls@nano.uni-hannover.de}
\author{U.~Zeitler}%
\author{R.~J.~Haug}
\affiliation{%
Institut f\"ur Festk\"orperphysik,
Universit\"at Hannover, Appelstr. 2, 30167 Hannover, Germany
}%


\begin{abstract}
\mbox{}\\[3ex] 
We have measured the temperature dependence 
of the conductivity $\sigma_{xx}$ of a two-dimensional electron system 
deep into the localized regime of the
quantum Hall plateau transition. Using variable-range hopping theory
we are able to extract directly the localization length $\xi$ from
this experiment. We use our results to
study the scaling behavior of $\xi$
as a function of the filling factor distance $|\delta \nu|$ to
the critical point of the transition.
We find for all samples a power-law behavior
\mbox{$\xi \propto |\delta\nu|^{-\gamma}$}  with a universal scaling exponent 
\mbox{$\gamma = 2.3$} as proposed theoretically.

\end{abstract}


\maketitle

\newlength{\plotwidth}          
\setlength{\plotwidth}{8.3cm}

The prominent feature of the quantum Hall effect (QHE) is the emergence
of quantized plateaus in the Hall resistance of a two-dimensional electron
system (2DES) 
reflecting the localization of states at the Fermi edge.
A deeper understanding of this fascinating phenomenon is gained
from regarding the transition region between
adjacent QHE plateaus~\cite{huckestein95,sondhi97}.
In this regime the dissipative transport is governed
by the delocalized  states in the vicinity of the Landau level center.
The degree of localization is expressed by the localization length $\xi$
denoting the typical extension of the electron wave function.
For any sample with a finite size $L$ theory predicts the
conductivity tensor in the plateau transition
to follow a scaling function $f\left(L/\xi\right)$ with
a diverging localization length 
\mbox{$\xi=|\delta\nu|^{-\gamma}$}~\cite{pruisken88}
where \mbox{$\delta\nu=\nu-\nu_c$} denotes the filling factor distance to
the critical point $\nu_c$ of the transition.
The critical exponent $\gamma=2.3$ was predicted
to be universal, its value determined
numerically~\cite{huckestein90} and validated
by variation of the sample size~\cite{koch91}.
Experimentally the scaling of the conductivity near the critical point
was verified with great success for different samples by temperature, current,
and frequency dependent measurements of the transition width~\cite{wei88,chow96,engel93}.
These new parameters introduce effective lengths
\mbox{$L_T\propto T^{p/2}$}, \mbox{$L_I\propto I^{p/(2+p)}$}, and
\mbox{$L_f\propto f^{1/z}$} and thereby add additional exponents $z$ and $p$.
These effective lengths replace the physical sample size $L$ in the
scaling functions $f(L/\xi)$.
Recent experiments extended the focus to the transition
from the quantum Hall state to the Hall
insulator~\cite{shahar95,shahar96,hilke98}.
In these investigations
striking similarities to the scaling behavior in the transition between
different QH states were observed~\cite{shahar97} hinting to
the same universality class for both types of
transitions~\cite{pan97,schaijk00}.

In spite of the great success of scaling theory in the QHE there are still
some experiments not fitting into the picture of universality.
Non-universal exponents were observed in the dependence of the
transition width on temperature~\cite{koch91b},
current~\cite{chow95} and frequency~\cite{hohls01a}.
Other experiments seem to contradict scaling theory at all,
both for the Hall plateau-insulator transition~\cite{shahar98} and the
transition between QHE plateaus~\cite{coleridge99,balaban98}.

However, before making conclusions on a general failure of scaling theory it
has to be considered that nearly all experiments
{\sl do not} measure the localization length $\xi$ directly.
Therefore an assumption about the functional form and exponents $z$ and
$p$ of the effective length $L_{\rm eff}(T,I,f)$ has to be made.
The non-universal scaling exponents deduced from
investigating the QHE transition width as a function of
$T$, $I$ and $f$  then only reflect non-universal exponents $p$ and $z$ in
$L_{\rm eff}$.
Additionally, the measurements mostly focus on the region close to critical points,
where $\xi$ becomes larger than $L_{\rm eff}$. In this regime, however,
the mechanism of electronic transport at nonzero temperature
or frequency is not thoroughly understood.
Addressing this problem Shimshoni found in a recent theoretical work 
for the Hall plateau-insulator transition 
that for $T>0$ quantum transport occurs only at some distance
to the critical point, namely in the regime of hopping conductivity~\cite{shimshoni99}.
Therefore, any lack of universal width scaling with $T$ or $f$ does not necessarily
allow to draw conclusions on the behavior of $\xi$.

In order to avoid such complications, we follow a different approach to scaling.
We directly evaluate the localization length in the well understood regime of
variable-range hopping (VRH) conductivity ~\cite{polyakov93}.
We have recently shown this method to be reliable for frequency dependent
measurements~\cite{hohls01b}.
VRH dominates the conductivity at low temperatures, when the localization
length becomes much smaller than the effective temperature length $L_T$.
In the QHE regime the VRH conductivity is given
as~\cite{ebert83,briggs83,polyakov93}
\begin{equation}
  \label{hoppingeq}
  \sigma_{xx}(T) = \sigma_0 \exp\left({-\sqrt{T_0/T}}\right),\quad
    k_BT_0=C\frac{e^2}{4\pi\epsilon\epsilon_0 \xi}~~,
\end{equation}
with a temperature dependent prefactor $\sigma_0\propto 1/T$.
The characteristic temperature $T_0$ is determined by the Coulomb energy
at a length scale given by the localization length $\xi$,
the dimensionless constant $C$ being in the order of unity.
With this direct access to $\xi\propto 1/T_0$ it is possible to test for
scaling behavior of $\xi$ at the edges of the plateau as long as we are
careful enough to stay within the localized regime.

Earlier experiments confirmed the expected temperature dependence of
$\sigma_{xx}(T)$
and extracted $\xi$ in the QHE-regime~\cite{koch95,furlan98},
but either did not analyze the scaling behavior or were restricted to a small
filling factor range close to the quasi-metallic regime. In a recent experiment
on the QH plateau-insulator transition VRH conductivity following
Eq.~\ref{hoppingeq} was also established to that regime and was used
to determine $\xi$, finding a rough agreement with the prediction of universal
scaling~\cite{murphy00}.

In this paper we show that when analyzing the conductivity deep into the
localized regime of the quantum Hall plateau transition
we find a clear universal scaling behavior
of the localization length \mbox{$\xi \propto |\delta\nu|^{-\gamma}$}
as function of the filling factor distance \mbox{$\delta\nu = \nu - \nu_c$}
to the critical point at $\nu_c$ with a universal \mbox{$\gamma = 2.3$}.
Such a universality is even observed in samples where
non-universal exponents extracted from
temperature dependent peak-width scaling are found.
Going even further, we show that the conductivities
$\sigma_{xx}(T,\delta\nu)$ in the VRH regime for \mbox{$\delta\nu < 0.3$}
can be scaled to  a
single parameter function of $|\delta\nu|^\gamma/T$ as
predicted by scaling theory for the critical regime.

\begin{table}
  \begin{tabular}[t]{cccccc} \hline \hline
 Sample & doping & $n_e$ & $\mu_e$ & $\nu_c$ & $\kappa$ \\
        &        & $10^{15}$ m$^{-2}$ & m$^2$/Vs & & \\ \hline
 S1 & $\delta$-Be & 2.1 & 2 & 1.29 & $\;\;0.66\pm 0.02$ \\
 S2 & $\delta$ Si & 3.2 & 4 & 1.62 & $\;\;0.60 \pm 0.02$ \\
 S3 & hom. Be & 2.4 & 12 & 1.52 & $\;\;0.62 \pm 0.03$ \\ \hline \hline
 \end{tabular}
 \caption{Sample characteristics: Type of extra doping, electron density
$n_e$ and mobility $\mu_e$ of the 2DES, the critical point
$\nu_c$~\cite{remark1},
and the width scaling exponent $\kappa$
for the \mbox{$\nu=1\rightarrow 2$} transition.}
\label{table}
\end{table}

The samples used in this work are based on modulation doped
GaAs/AlGaAs heterostructures with additional scatterers in
the active region of the 2DES \cite{ploog87}.  The scatterers
are provided by doping the GaAs close to the heterojunction
with Si or Be, either as a $\delta$-layer or a weak homogeneous background.
This results in relatively low mobilities of a few m$^2$/Vs
(see table~\ref{table}).
In order to allow a highly sensitive
two-point measurement of very low conductivities
the samples were patterned into Corbino geometry
using contacts fabricated by standard Ni/Au/Ge alloy annealing.
Here the conductivity is given by
\mbox{$\sigma_{xx} = \left(I/2\pi V\right)\ln\left(r_2/r_1\right)$}
where \mbox{$r_1 = 500\,\mu$m} and \mbox{$r_2 = 550\,\mu$m} are the inner and
the outer radius of the Corbino disk.
The samples were mounted onto the cold finger of a
dilution refrigerator with base temperature below 20~mK and positioned into
the center of a superconducting solenoid.

The sample conductivity $\sigma_{xx}(B,T)$
as a function of magnetic field $B$ and temperature $T$
was extracted from individually measured $I$-$V$ characteristics
by numerical differentiation in the linear regime close to
zero bias. For each curve we identified the linear regime
which ranged from a few $\mu$V around the QHE plateau transition at
low temperatures up to a few mV in the plateau center.
This method allowed us to measure the conductivity accurately
in a huge range between $10^{-13}~\Omega^{-1}$ in the plateau center
up to $10^{-5}~\Omega^{-1}$ in the maximum of the plateau transition.

\begin{figure}
  \begin{center}
  \resizebox{\plotwidth}{!}{\includegraphics{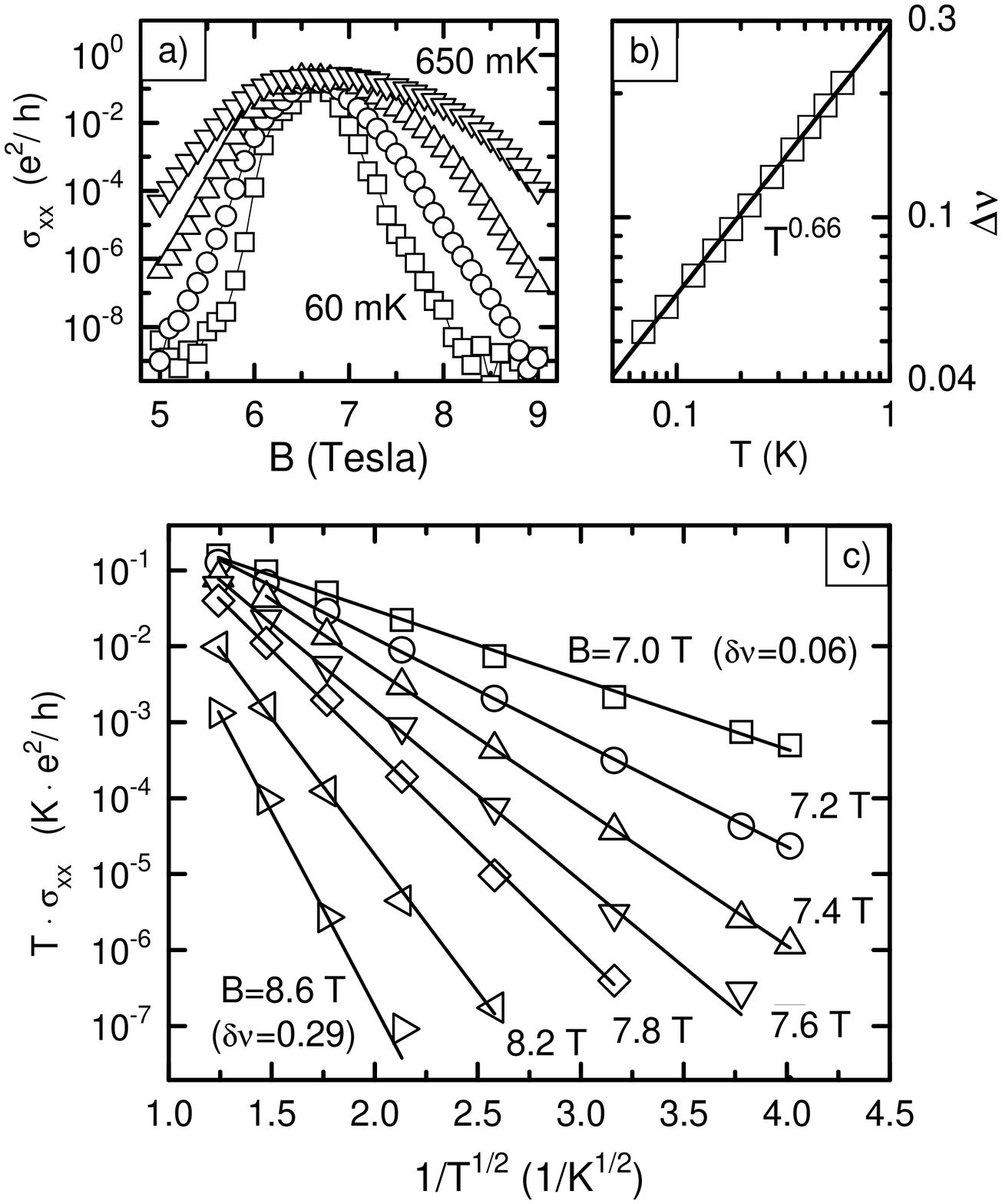}}
  \end{center}
  \caption{
     a)~Conductivity $\sigma_{xx}$
        of sample S1 at different temperatures for the
        transition between plateaus $\nu=2$ and $\nu=1$.
     b)~Full width at half maximum $\Delta \nu$ of the conductivity
        peak fitted by a power law \mbox{$\Delta\nu\propto T^{\kappa}$}.
     c)~Variable-range hopping fit of $\sigma_{xx}$ to
        Eq.~\ref{hoppingeq} with prefactor \mbox{$\sigma_0 \propto 1/T$}.
        The axes are rescaled to show straight line for the prediction of
        Eq.~\ref{hoppingeq}.}
  \label{data}
\end{figure}

In Fig.~\ref{data}a $\sigma_{xx}(B,T)$ of sample S1 is shown for the transition
between the QHE plateaus at \mbox{$\nu=1$} and \mbox{$\nu=2$}.
We concentrate our studies on this transition where the
spin gap is sufficiently large to exclude activated transport.
The critical point $\nu_c$ of this transition, listed in 
table~\ref{table}, is given by the position
of the maximum in $\sigma_{xx}(\nu)$.

Before analyzing the VRH transport  we tested the
conventional scaling behavior of the plateau transition width.
Data for the full-width at half maximum $\Delta \nu$ of the peak in
$\sigma_{xx}$ as a function of filling factor \mbox{$\nu = hn/eB$}
are shown in Fig.~\ref{data}b for sample S1.
For all investigated samples $\Delta \nu$ follows
a power-law behavior \mbox{$\Delta\nu\propto T^\kappa$}
down to temperatures below 60~mK.
However, as already
reported in earlier work on similar samples~\cite{koch91b}
the critical exponents $\kappa$ as presented
in table~\ref{table} considerably
deviate from the proposed universal value \mbox{$\kappa = 0.43$}~\cite{wei88}.

Let us now turn to a closer analysis of the VRH regime.
As shown in Fig.~\ref{data}c for sample S1 the data fit well the
predicted temperature behavior of Eq.~\ref{hoppingeq} with a prefactor
\mbox{$\sigma_0\propto 1/T$}.
We also tested for activated behavior, Coulomb gap with constant $\sigma_0$,
and Mott hopping 
\mbox{$\sigma_{xx}\propto  T^{-m}\exp\left((T_0/T)^{1/3}\right)$} for
various $m$, all matching our data worse.
From these fits to Eq.~\ref{hoppingeq}
we are able to extract the characteristic
temperature $T_0$ in the VRH regime. To stay well inside this
localized regime we only take into account filling factors where the low
temperature conductivity is at least two orders of magnitude below
the critical conductivity $\sigma_c$.

\begin{figure}
  \begin{center}
  \resizebox{0.8\plotwidth}{!}{\includegraphics{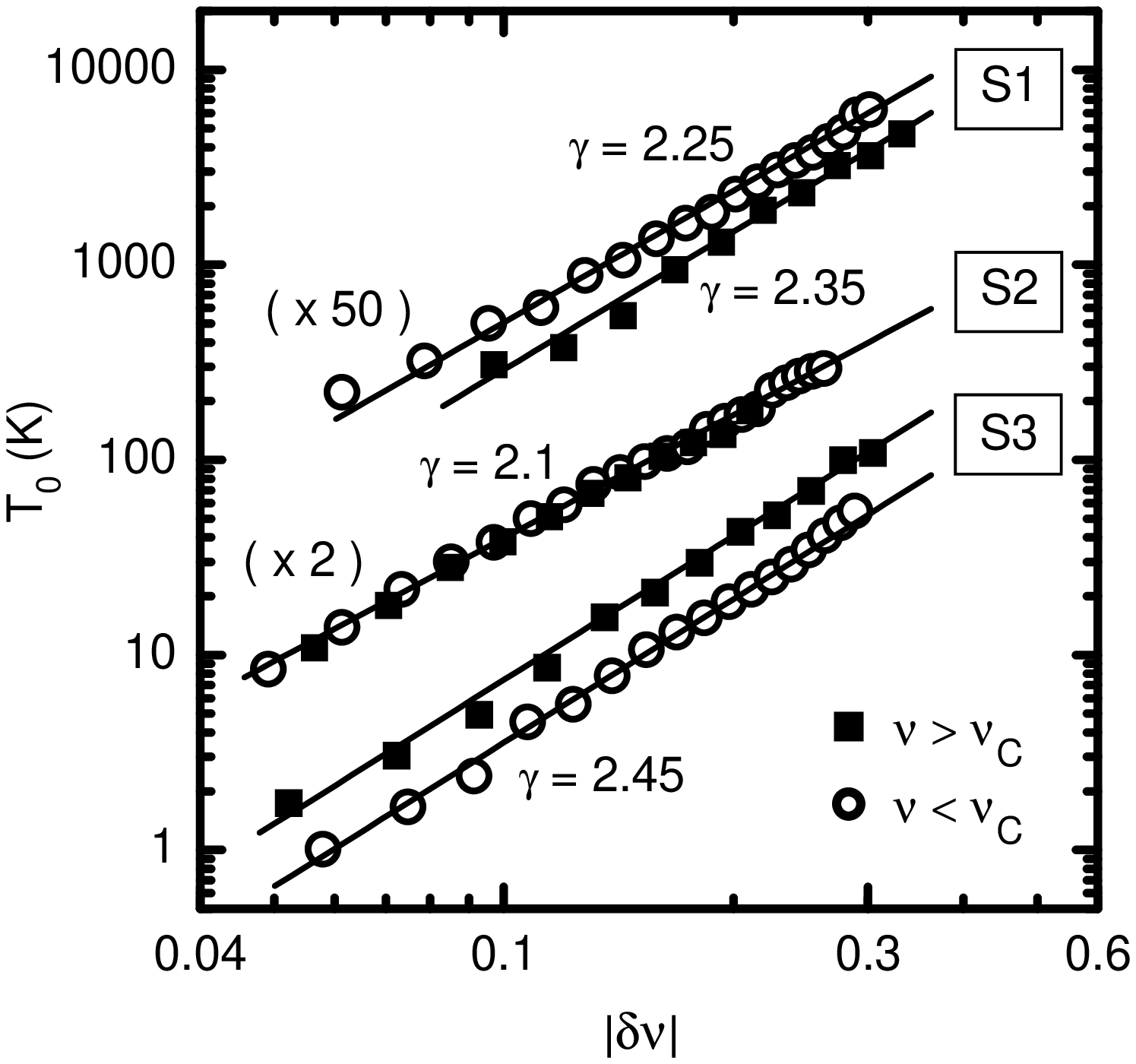}}
  \end{center}
  \caption{Characteristic temperature $T_0$ as function of the distance
    $|\delta\nu|$ to the critical point for both sides
    of the \mbox{$\nu=1\rightarrow 2$}
    plateau transition on a logarithmic scale.
    The data for different samples S1, S2 and S3 are shifted for clarity.
    The lines show power-law fits \mbox{$T_0 \propto |\delta\nu|^\gamma$}.
    Within the 2$\sigma$-confidence interval of $\pm0.2$ the extracted
    exponents agree with the expected \mbox{$\gamma=2.3$}.
    }
  \label{hoppingT}
\end{figure}

In Fig.~\ref{hoppingT} the characteristic temperature $T_0$ is plotted
against the distance \mbox{$|\delta\nu|=|\nu-\nu_c|$}
to the critical point $\nu_c$. 
The data follows a power-law behavior
\mbox{$T_0 \propto |\delta\nu|^\gamma$}  up to distances as large
as \mbox{$|\delta\nu|\approx 0.3$}.
This demonstrates that the localization length \mbox{$\xi\propto 1/T_0$}
follows a scaling behavior deep into the localized regime. 

The most interesting result is that we find an accurate universal
scaling of the localization length $\xi$ with temperature.
Our extracted values for $\gamma$ agree within experimental accuracy
well with the numerical results
\mbox{$\gamma = 2.3$} (\cite{huckestein95} and references therein) as well as
with the exponents determined from size dependent scaling
experiments~\cite{koch91}. This underlines the fact that our direct
scaling analysis in the VRH regime suits much better as an
access to the localization length and its scaling behavior
than temperature dependent peak-width scaling where no universality
is observed. In fact, for
such a conventional analysis we are still lacking a sufficient
knowledge of the temperature dependent transport mechanisms
in the metallic regime of the QHE plateau transition.

Until now we used the term scaling in a rather reduced sense as synonym for
a power-law behavior of the localization length $\xi$.
Of course scaling includes much more,
namely the existence of a single parameter scaling function
\mbox{$\sigma_{xx}=f(x)$} with a parameter $x(T,\delta\nu)$.
Using our above results in the VRH regime an appropriate definition
is \mbox{$x=|\delta\nu|^\gamma/T \propto T_0/T$} with the above deduced
exponents $\gamma$. The postulation of
single parameter scaling together with the finding of 
\mbox{$\sigma_0\propto 1/T$} then fixes the prefactor in Eq.~\ref{hoppingeq} to
\mbox{$\sigma_0(\delta\nu,T)= \sigma^\ast|\delta\nu|^\gamma/T$}
with a constant $\sigma^\ast$. This yields a scaling
of the conductivity in the form
\begin{equation}
  \sigma_{xx}\left(\frac{|\delta\nu|^\gamma}{T}\right) = \sigma^\ast
  \frac{|\delta\nu|^\gamma}{T} \exp\left(-\sqrt{T^\ast\frac{|\delta\nu|^\gamma}{T}}\right)
  \label{rescaleeq}
\end{equation}
where $T^\ast$ is constant.

\begin{figure}
  \begin{center}
  \resizebox{\plotwidth}{!}{\includegraphics{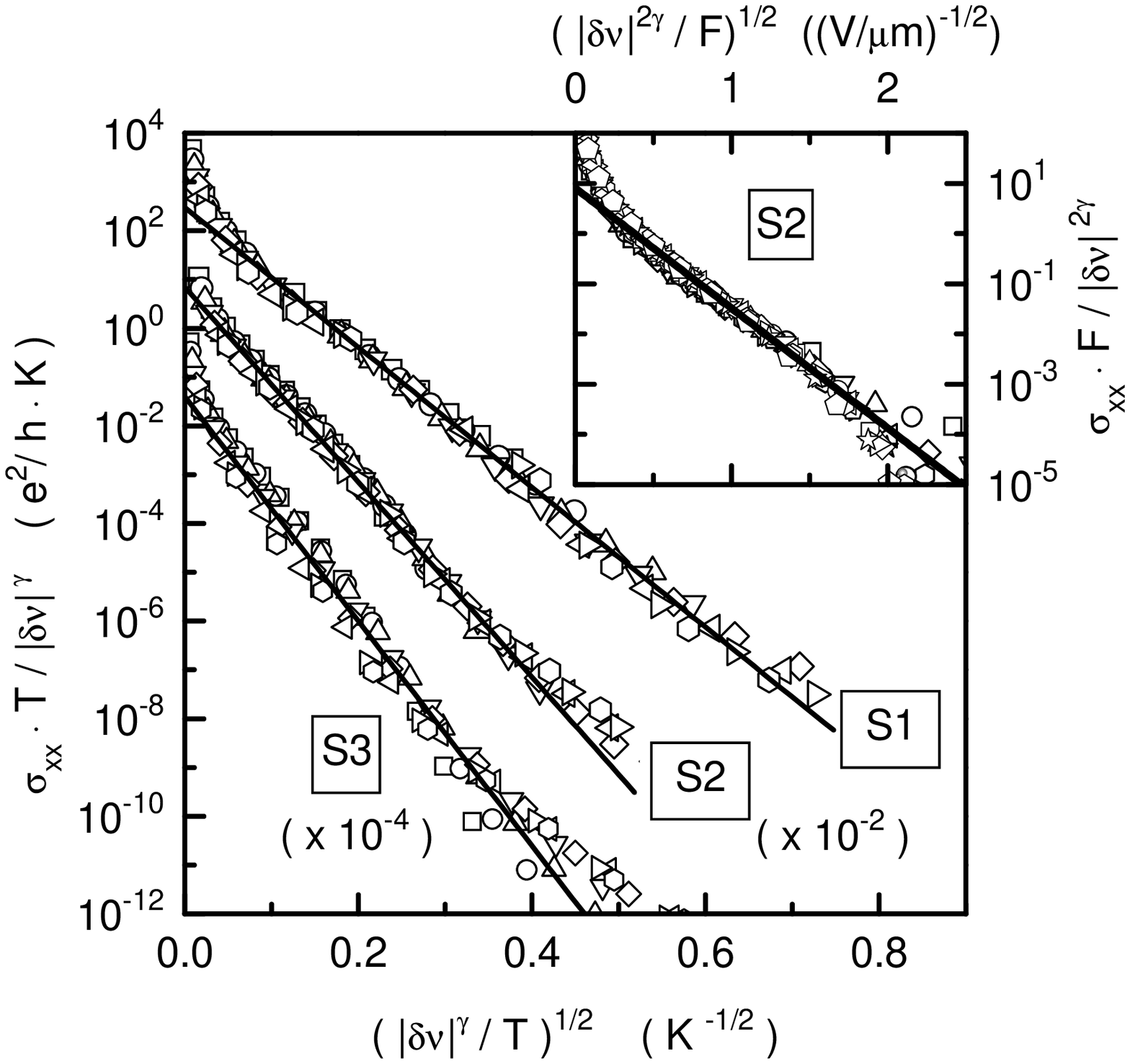}}
  \end{center}
  \caption{Rescaling of the conductivity $\sigma_{xx}\left(T,\nu\right)$,
    presented for sample S1 in figure \ref{data}c),
    as function of a single parameter \mbox{$x=|\delta\nu|^{\gamma}/T$} for
    \mbox{$\nu>\nu_c$} ($T=60\ldots 700$~mK, \mbox{$|\delta\nu| < 0.3$}).
    Inset: Test of scaling with a single parameter
    \mbox{$y=|\delta\nu|^{2\gamma}/F$} for the dependence on electric field
    (\mbox{$F = 2\ldots 200$~V/m}, \mbox{$\sigma_{xx}\geq 10^{-4}e^2/h$}).
}
  \label{rescale}
\end{figure}

In Fig.~\ref{rescale} we have plotted all the conductivity data
$\sigma_{xx}\left(\nu,T\right)$ for \mbox{$\nu > \nu_c$} as
function of a single parameter $|\delta\nu|^\gamma/T$.
The values for $\gamma$ are taken from the power-law fits of $T_0$
as shown in Fig.~\ref{hoppingT}. Rescaling the axes in a proper way
indeed shows that all experimental points fall onto
straight lines as represented by the scaling function in
Eq.~\ref{rescaleeq}. Therefore, the conductivity in the VRH
regime for \mbox{$\delta\nu < 0.3$} and \mbox{$\nu > \nu_c$}
follows the stringent postulation of single parameter scaling and
thus demonstrates the large validity range of
the universal scaling phenomenon in the transition between QHE plateaus.

For \mbox{$\nu<\nu_c$}, i.e. approaching the spin gap induced plateau at 
\mbox{$\nu=1$},
the single parameter rescaling does not look as good as for approaching the
Landau gap at \mbox{$\nu=2$}, although Fig.~\ref{hoppingT} showed the same
quality of $T_0$-scaling for both sides of the transition. This discrepancy
probably mirrors a more complicated behavior of the prefactor $\sigma_0$ in
Eq.~\ref{hoppingeq}. Obviously it can no more be written as a simple function
\mbox{$|\delta\nu|^\gamma/T \propto T_0/T$}. Such a more complicated
behavior in this regime is no surprise taking into account the
redistribution of electrons between spin states and the emerging of spin
textures as a function of $\nu$ at this filling range~\cite{girvin97}.

In addition to the temperature dependence of $\sigma_{xx}$
in the VRH regime we have also investigated its dependence
on the electric field strength $F$.
In the nonlinear regime at high voltages the effect of $F$
can be interpreted as an effective temperature
\mbox{$T_F = eF\xi/2k_B$}~\cite{polyakov93}.
Using Eq.~\ref{hoppingeq} the conductivity
is then rewritten to
\mbox{$\sigma_{xx}=\sigma^F_0(F)\exp\left(-\sqrt{F_0/F}\right)$}
with \mbox{$F_0\propto T_0/\xi \propto 1/\xi^2$}. 
Analogous to the single parameter
scaling with temperature in the VRH regime the natural reduced parameter
for scaling with electric field is
defined by \mbox{$y(F,\nu)=|\delta\nu|^{2\gamma}/F$}.
Applying such an analysis to our data without any additional fit parameter
(i.e. using the exponents  $\gamma$ as determined from $T_0(\nu)$,
see Fig.~\ref{hoppingT})
we find for \mbox{$\nu > \nu_c$} for all data in the range  
\mbox{$\sigma_{xx} \geq 10^{-4}e^2/h$}
a single parameter scaling; an example for sample S2 is shown 
in the inset of figure~\ref{rescale}.

In conclusion we have investigated the temperature and electric field
dependence of the conductivity in the QHE plateau transition for samples
where no universal behavior is found in
the conventional temperature dependent peak-width scaling experiments.
We find a revival of universality in the VRH regime where the
localization length $\xi$ scales as \mbox{$\xi \propto |\delta \nu|^{-\gamma}$}
with an experimentally deduced scaling exponent
close to the theoretically expected value  \mbox{$\gamma=2.3$}.
Even further, we have shown that all the data on
the Landau gap side of the transition can be rescaled on a single parameter
function over more than five orders of magnitude in the conductivity.

We thank F. Evers, B. Huckestein, B. Kramer and D. G. Polyakov for
stimulating discussions. The samples were grown by K. Ploog at the
Max Planck-Institut f\"ur Festk\"orperforschung.\\


\begin{thebibliography}{32}
\expandafter\ifx\csname natexlab\endcsname\relax\def\natexlab#1{#1}\fi
\expandafter\ifx\csname bibnamefont\endcsname\relax
  \def\bibnamefont#1{#1}\fi
\expandafter\ifx\csname bibfnamefont\endcsname\relax
  \def\bibfnamefont#1{#1}\fi
\expandafter\ifx\csname citenamefont\endcsname\relax
  \def\citenamefont#1{#1}\fi
\expandafter\ifx\csname url\endcsname\relax
  \def\url#1{\texttt{#1}}\fi
\expandafter\ifx\csname urlprefix\endcsname\relax\def\urlprefix{URL }\fi
\providecommand{\bibinfo}[2]{#2}
\providecommand{\eprint}[2][]{\url{#2}}

\bibitem[{\citenamefont{Huckestein}(1995)}]{huckestein95}
\bibinfo{author}{\bibfnamefont{B.}~\bibnamefont{Huckestein}},
  \bibinfo{journal}{Rev. Mod. Phys.}
  \textbf{\bibinfo{volume}{67}}(\bibinfo{number}{2}), \bibinfo{pages}{357}
  (\bibinfo{year}{1995}).

\bibitem[{\citenamefont{Sondhi et~al.}(1997)\citenamefont{Sondhi, Girvin,
  Carini, and Shahar}}]{sondhi97}
\bibinfo{author}{\bibfnamefont{S.~L.} \bibnamefont{Sondhi}},
  \bibinfo{author}{\bibfnamefont{S.~M.} \bibnamefont{Girvin}},
  \bibinfo{author}{\bibfnamefont{J.~P.} \bibnamefont{Carini}},
  \bibnamefont{and} \bibinfo{author}{\bibfnamefont{D.}~\bibnamefont{Shahar}},
  \bibinfo{journal}{Rev. Mod. Phys.}
  \textbf{\bibinfo{volume}{69}}(\bibinfo{number}{1}), \bibinfo{pages}{315}
  (\bibinfo{year}{1997}).

\bibitem[{\citenamefont{Pruisken}(1988)}]{pruisken88}
\bibinfo{author}{\bibfnamefont{A.~M.~M.} \bibnamefont{Pruisken}},
  \bibinfo{journal}{Phys. Rev. Lett.}
  \textbf{\bibinfo{volume}{61}}(\bibinfo{number}{11}), \bibinfo{pages}{1297}
  (\bibinfo{year}{1988}).

\bibitem[{\citenamefont{Huckestein and Kramer}(1990)}]{huckestein90}
\bibinfo{author}{\bibfnamefont{B.}~\bibnamefont{Huckestein}} \bibnamefont{and}
  \bibinfo{author}{\bibfnamefont{B.}~\bibnamefont{Kramer}},
  \bibinfo{journal}{Phys. Rev. Lett.}
  \textbf{\bibinfo{volume}{64}}(\bibinfo{number}{12}), \bibinfo{pages}{1437}
  (\bibinfo{year}{1990}).

\bibitem[{\citenamefont{Koch et~al.}(1991{\natexlab{a}})\citenamefont{Koch,
  Haug, v.~Klitzing, and Ploog}}]{koch91}
\bibinfo{author}{\bibfnamefont{S.}~\bibnamefont{Koch}},
  \bibinfo{author}{\bibfnamefont{R.~J.} \bibnamefont{Haug}},
  \bibinfo{author}{\bibfnamefont{K.}~\bibnamefont{v.~Klitzing}},
  \bibnamefont{and} \bibinfo{author}{\bibfnamefont{K.}~\bibnamefont{Ploog}},
  \bibinfo{journal}{Phys. Rev. Lett.}
  \textbf{\bibinfo{volume}{67}}(\bibinfo{number}{7}), \bibinfo{pages}{883}
  (\bibinfo{year}{1991}{\natexlab{a}}).

\bibitem[{\citenamefont{Wei et~al.}(1988)\citenamefont{Wei, Tsui, Paalanen, and
  Pruisken}}]{wei88}
\bibinfo{author}{\bibfnamefont{H.~P.} \bibnamefont{Wei}},
  \bibinfo{author}{\bibfnamefont{D.~C.} \bibnamefont{Tsui}},
  \bibinfo{author}{\bibfnamefont{M.~A.} \bibnamefont{Paalanen}},
  \bibnamefont{and} \bibinfo{author}{\bibfnamefont{A.~M.~M.}
  \bibnamefont{Pruisken}}, \bibinfo{journal}{Phys. Rev. Lett.}
  \textbf{\bibinfo{volume}{61}}(\bibinfo{number}{11}), \bibinfo{pages}{1294}
  (\bibinfo{year}{1988}).

\bibitem[{\citenamefont{Chow et~al.}(1996)\citenamefont{Chow, Wei, Girvin, and
  Shayegan}}]{chow96}
\bibinfo{author}{\bibfnamefont{E.}~\bibnamefont{Chow}},
  \bibinfo{author}{\bibfnamefont{H.~P.} \bibnamefont{Wei}},
  \bibinfo{author}{\bibfnamefont{S.~M.} \bibnamefont{Girvin}},
  \bibnamefont{and} \bibinfo{author}{\bibfnamefont{M.}~\bibnamefont{Shayegan}},
  \bibinfo{journal}{Phys. Rev. Lett.}
  \textbf{\bibinfo{volume}{77}}(\bibinfo{number}{6}), \bibinfo{pages}{1143}
  (\bibinfo{year}{1996}).

\bibitem[{\citenamefont{Engel et~al.}(1993)\citenamefont{Engel, Shahar, Kurdak,
  and Tsui}}]{engel93}
\bibinfo{author}{\bibfnamefont{L.~W.} \bibnamefont{Engel}},
  \bibinfo{author}{\bibfnamefont{D.}~\bibnamefont{Shahar}},
  \bibinfo{author}{\bibfnamefont{C.}~\bibnamefont{Kurdak}}, \bibnamefont{and}
  \bibinfo{author}{\bibfnamefont{D.~C.} \bibnamefont{Tsui}},
  \bibinfo{journal}{Phys. Rev. Lett.}
  \textbf{\bibinfo{volume}{71}}(\bibinfo{number}{16}), \bibinfo{pages}{2638}
  (\bibinfo{year}{1993}).

\bibitem[{\citenamefont{Shahar et~al.}(1995)\citenamefont{Shahar, Tsui,
  Shayegan, Bhatt, and Cunningham}}]{shahar95}
\bibinfo{author}{\bibfnamefont{D.}~\bibnamefont{Shahar}},
  \bibinfo{author}{\bibfnamefont{D.~C.} \bibnamefont{Tsui}},
  \bibinfo{author}{\bibfnamefont{M.}~\bibnamefont{Shayegan}},
  \bibinfo{author}{\bibfnamefont{R.~N.} \bibnamefont{Bhatt}}, \bibnamefont{and}
  \bibinfo{author}{\bibfnamefont{J.~E.} \bibnamefont{Cunningham}},
  \bibinfo{journal}{Phys. Rev. Lett.}
  \textbf{\bibinfo{volume}{74}}(\bibinfo{number}{22}), \bibinfo{pages}{4511}
  (\bibinfo{year}{1995}).

\bibitem[{\citenamefont{Shahar et~al.}(1996)\citenamefont{Shahar, Tsui,
  Shayegan, Shimshoni, and Sondhi}}]{shahar96}
\bibinfo{author}{\bibfnamefont{D.}~\bibnamefont{Shahar}},
  \bibinfo{author}{\bibfnamefont{D.~C.} \bibnamefont{Tsui}},
  \bibinfo{author}{\bibfnamefont{M.}~\bibnamefont{Shayegan}},
  \bibinfo{author}{\bibfnamefont{E.}~\bibnamefont{Shimshoni}},
  \bibnamefont{and} \bibinfo{author}{\bibfnamefont{S.~L.}
  \bibnamefont{Sondhi}}, \bibinfo{journal}{Science}
  \textbf{\bibinfo{volume}{274}}, \bibinfo{pages}{589} (\bibinfo{year}{1996}).

\bibitem[{\citenamefont{Hilke et~al.}(1998)\citenamefont{Hilke, Shahar, Song,
  Tsui, Xie, and Monroe}}]{hilke98}
\bibinfo{author}{\bibfnamefont{M.}~\bibnamefont{Hilke}},
  \bibinfo{author}{\bibfnamefont{D.}~\bibnamefont{Shahar}},
  \bibinfo{author}{\bibfnamefont{S.~H.} \bibnamefont{Song}},
  \bibinfo{author}{\bibfnamefont{D.~C.} \bibnamefont{Tsui}},
  \bibinfo{author}{\bibfnamefont{Y.~H.} \bibnamefont{Xie}}, \bibnamefont{and}
  \bibinfo{author}{\bibfnamefont{D.}~\bibnamefont{Monroe}},
  \bibinfo{journal}{Nature} \textbf{\bibinfo{volume}{395}},
  \bibinfo{pages}{675} (\bibinfo{year}{1998}).

\bibitem[{\citenamefont{Shahar et~al.}(1997)\citenamefont{Shahar, Tsui,
  Shayegan, Shimshoni, and Sondhi}}]{shahar97}
\bibinfo{author}{\bibfnamefont{D.}~\bibnamefont{Shahar}},
  \bibinfo{author}{\bibfnamefont{D.~C.} \bibnamefont{Tsui}},
  \bibinfo{author}{\bibfnamefont{M.}~\bibnamefont{Shayegan}},
  \bibinfo{author}{\bibfnamefont{E.}~\bibnamefont{Shimshoni}},
  \bibnamefont{and} \bibinfo{author}{\bibfnamefont{S.~L.}
  \bibnamefont{Sondhi}}, \bibinfo{journal}{Phys. Rev. Lett.}
  \textbf{\bibinfo{volume}{79}}(\bibinfo{number}{3}), \bibinfo{pages}{479}
  (\bibinfo{year}{1997}).

\bibitem[{\citenamefont{Pan et~al.}(1997)\citenamefont{Pan, Shahar, Tsui, Wei,
  and Razeghi}}]{pan97}
\bibinfo{author}{\bibfnamefont{W.}~\bibnamefont{Pan}},
  \bibinfo{author}{\bibfnamefont{D.}~\bibnamefont{Shahar}},
  \bibinfo{author}{\bibfnamefont{D.~C.} \bibnamefont{Tsui}},
  \bibinfo{author}{\bibfnamefont{H.~P.} \bibnamefont{Wei}}, \bibnamefont{and}
  \bibinfo{author}{\bibfnamefont{M.}~\bibnamefont{Razeghi}},
  \bibinfo{journal}{Phys. Rev. B}
  \textbf{\bibinfo{volume}{55}}(\bibinfo{number}{23}), \bibinfo{pages}{15431}
  (\bibinfo{year}{1997}).

\bibitem[{\citenamefont{{van~Schaijk} et~al.}(2000)\citenamefont{{van~Schaijk},
  {de~Visser}, Olsthoorn, Wei, and Pruisken}}]{schaijk00}
\bibinfo{author}{\bibfnamefont{R.~T.~F.} \bibnamefont{{van~Schaijk}}},
  \bibinfo{author}{\bibfnamefont{A.}~\bibnamefont{{de~Visser}}},
  \bibinfo{author}{\bibfnamefont{S.}~\bibnamefont{Olsthoorn}},
  \bibinfo{author}{\bibfnamefont{H.~P.} \bibnamefont{Wei}}, \bibnamefont{and}
  \bibinfo{author}{\bibfnamefont{A.~M.~M.} \bibnamefont{Pruisken}},
  \bibinfo{journal}{Phys. Rev. Lett.} \textbf{\bibinfo{volume}{84}},
  \bibinfo{pages}{1567} (\bibinfo{year}{2000}).

\bibitem[{\citenamefont{Koch et~al.}(1991{\natexlab{b}})\citenamefont{Koch,
  Haug, v.~Klitzing, and Ploog}}]{koch91b}
\bibinfo{author}{\bibfnamefont{S.}~\bibnamefont{Koch}},
  \bibinfo{author}{\bibfnamefont{R.~J.} \bibnamefont{Haug}},
  \bibinfo{author}{\bibfnamefont{K.}~\bibnamefont{v.~Klitzing}},
  \bibnamefont{and} \bibinfo{author}{\bibfnamefont{K.}~\bibnamefont{Ploog}},
  \bibinfo{journal}{Phys. Rev. B}
  \textbf{\bibinfo{volume}{43}}(\bibinfo{number}{8}), \bibinfo{pages}{6828}
  (\bibinfo{year}{1991}{\natexlab{b}}).

\bibitem[{\citenamefont{Chow and Wei}(1995)}]{chow95}
\bibinfo{author}{\bibfnamefont{E.}~\bibnamefont{Chow}} \bibnamefont{and}
  \bibinfo{author}{\bibfnamefont{H.~P.} \bibnamefont{Wei}},
  \bibinfo{journal}{Phys. Rev. B}
  \textbf{\bibinfo{volume}{52}}(\bibinfo{number}{19}), \bibinfo{pages}{13749}
  (\bibinfo{year}{1995}).

\bibitem[{\citenamefont{Hohls et~al.}(2001{\natexlab{a}})\citenamefont{Hohls,
  Zeitler, Haug, and Pierz}}]{hohls01a}
\bibinfo{author}{\bibfnamefont{F.}~\bibnamefont{Hohls}},
  \bibinfo{author}{\bibfnamefont{U.}~\bibnamefont{Zeitler}},
  \bibinfo{author}{\bibfnamefont{R.~J.} \bibnamefont{Haug}}, \bibnamefont{and}
  \bibinfo{author}{\bibfnamefont{K.}~\bibnamefont{Pierz}},
  \bibinfo{journal}{Physica B}
  \textbf{\bibinfo{volume}{298}}(\bibinfo{number}{1-4}), \bibinfo{pages}{88}
  (\bibinfo{year}{2001}{\natexlab{a}}).

\bibitem[{\citenamefont{Shahar et~al.}(1998)\citenamefont{Shahar, Hilke, Li,
  Tsui, Sondhi, and Razeghi}}]{shahar98}
\bibinfo{author}{\bibfnamefont{D.}~\bibnamefont{Shahar}},
  \bibinfo{author}{\bibfnamefont{M.}~\bibnamefont{Hilke}},
  \bibinfo{author}{\bibfnamefont{C.~C.} \bibnamefont{Li}},
  \bibinfo{author}{\bibfnamefont{D.~C.} \bibnamefont{Tsui}},
  \bibinfo{author}{\bibfnamefont{S.~L.} \bibnamefont{Sondhi}},
  \bibnamefont{and} \bibinfo{author}{\bibfnamefont{M.}~\bibnamefont{Razeghi}},
  \bibinfo{journal}{Solid State Commun.}
  \textbf{\bibinfo{volume}{107}}(\bibinfo{number}{1}), \bibinfo{pages}{19}
  (\bibinfo{year}{1998}).

\bibitem[{\citenamefont{Balaban et~al.}(1998)\citenamefont{Balaban, Meirav, and
  {Bar-Joseph}}}]{balaban98}
\bibinfo{author}{\bibfnamefont{N.~Q.} \bibnamefont{Balaban}},
  \bibinfo{author}{\bibfnamefont{U.}~\bibnamefont{Meirav}}, \bibnamefont{and}
  \bibinfo{author}{\bibfnamefont{I.}~\bibnamefont{{Bar-Joseph}}},
  \bibinfo{journal}{Phys. Rev. Lett.}
  \textbf{\bibinfo{volume}{81}}(\bibinfo{number}{22}), \bibinfo{pages}{4967}
  (\bibinfo{year}{1998}).

\bibitem[{\citenamefont{Coleridge}(1999)}]{coleridge99}
\bibinfo{author}{\bibfnamefont{P.~T.} \bibnamefont{Coleridge}},
  \bibinfo{journal}{Phys. Rev. B}
  \textbf{\bibinfo{volume}{60}}(\bibinfo{number}{7}), \bibinfo{pages}{4493}
  (\bibinfo{year}{1999}).

\bibitem[{\citenamefont{Shimshoni}(1999)}]{shimshoni99}
\bibinfo{author}{\bibfnamefont{E.}~\bibnamefont{Shimshoni}},
  \bibinfo{journal}{Phys. Rev. B}
  \textbf{\bibinfo{volume}{60}}(\bibinfo{number}{15}), \bibinfo{pages}{10691}
  (\bibinfo{year}{1999}).

\bibitem[{\citenamefont{Polyakov and
  Shklovskii}(1993{\natexlab{a}})}]{polyakov93}
\bibinfo{author}{\bibfnamefont{D.~G.} \bibnamefont{Polyakov}} \bibnamefont{and}
  \bibinfo{author}{\bibfnamefont{B.~I.} \bibnamefont{Shklovskii}},
  \bibinfo{journal}{Phys. Rev. B}
  \textbf{\bibinfo{volume}{48}}(\bibinfo{number}{15}), \bibinfo{pages}{11167}
  (\bibinfo{year}{1993}{\natexlab{b}});
  \bibinfo{journal}{Phys. Rev. Lett.}
  \textbf{\bibinfo{volume}{70}}(\bibinfo{number}{24}), \bibinfo{pages}{3796}
  (\bibinfo{year}{1993}{\natexlab{a}}).

\bibitem[{\citenamefont{Hohls et~al.}(2001{\natexlab{b}})\citenamefont{Hohls,
  Zeitler, and Haug}}]{hohls01b}
\bibinfo{author}{\bibfnamefont{F.}~\bibnamefont{Hohls}},
  \bibinfo{author}{\bibfnamefont{U.}~\bibnamefont{Zeitler}}, \bibnamefont{and}
  \bibinfo{author}{\bibfnamefont{R.~J.} \bibnamefont{Haug}},
  \bibinfo{journal}{Phys. Rev. Lett.}
  \textbf{\bibinfo{volume}{86}}(\bibinfo{number}{122}), \bibinfo{pages}{5124}
  (\bibinfo{year}{2001}{\natexlab{b}}).

\bibitem[{\citenamefont{Ebert et~al.}(1983)\citenamefont{Ebert, von Klitzing,
  Probst, Schuberth, Ploog, and Weimann}}]{ebert83}
\bibinfo{author}{\bibfnamefont{G.}~\bibnamefont{Ebert}},
  \bibinfo{author}{\bibfnamefont{K.}~\bibnamefont{von Klitzing}},
  \bibinfo{author}{\bibfnamefont{C.}~\bibnamefont{Probst}},
  \bibinfo{author}{\bibfnamefont{E.}~\bibnamefont{Schuberth}},
  \bibinfo{author}{\bibfnamefont{K.}~\bibnamefont{Ploog}}, \bibnamefont{and}
  \bibinfo{author}{\bibfnamefont{G.}~\bibnamefont{Weimann}},
  \bibinfo{journal}{Solid State Commun.} \textbf{\bibinfo{volume}{45}},
  \bibinfo{pages}{625} (\bibinfo{year}{1983}).

\bibitem[{\citenamefont{Briggs et~al.}(1983)\citenamefont{Briggs, Guldner,
  Vieren, Voos, Hirtz, and Razeghi}}]{briggs83}
\bibinfo{author}{\bibfnamefont{A.}~\bibnamefont{Briggs}},
  \bibinfo{author}{\bibfnamefont{Y.}~\bibnamefont{Guldner}},
  \bibinfo{author}{\bibfnamefont{J.~P.} \bibnamefont{Vieren}},
  \bibinfo{author}{\bibfnamefont{M.}~\bibnamefont{Voos}},
  \bibinfo{author}{\bibfnamefont{J.~P.} \bibnamefont{Hirtz}}, \bibnamefont{and}
  \bibinfo{author}{\bibfnamefont{M.}~\bibnamefont{Razeghi}},
  \bibinfo{journal}{Phys. Rev. B} \textbf{\bibinfo{volume}{27}},
  \bibinfo{pages}{6549} (\bibinfo{year}{1983}).

\bibitem[{\citenamefont{Koch et~al.}(1995)\citenamefont{Koch, Haug,
  v.~Klitzing, and Ploog}}]{koch95}
\bibinfo{author}{\bibfnamefont{S.}~\bibnamefont{Koch}},
  \bibinfo{author}{\bibfnamefont{R.~J.} \bibnamefont{Haug}},
  \bibinfo{author}{\bibfnamefont{K.}~\bibnamefont{v.~Klitzing}},
  \bibnamefont{and} \bibinfo{author}{\bibfnamefont{K.}~\bibnamefont{Ploog}},
  \bibinfo{journal}{Semicond. Sci. Technol.} \textbf{\bibinfo{volume}{10}},
  \bibinfo{pages}{209} (\bibinfo{year}{1995}).

\bibitem[{\citenamefont{Furlan}(1998)}]{furlan98}
\bibinfo{author}{\bibfnamefont{M.}~\bibnamefont{Furlan}},
  \bibinfo{journal}{Phys. Rev. B}
  \textbf{\bibinfo{volume}{57}}(\bibinfo{number}{23}), \bibinfo{pages}{14818}
  (\bibinfo{year}{1998}).

\bibitem[{\citenamefont{Murphy et~al.}(2000)\citenamefont{Murphy, Hicks, Liu,
  Chung, Goldammer, and Santos}}]{murphy00}
\bibinfo{author}{\bibfnamefont{S.~Q.} \bibnamefont{Murphy}},
  \bibinfo{author}{\bibfnamefont{J.~L.} \bibnamefont{Hicks}},
  \bibinfo{author}{\bibfnamefont{W.~K.} \bibnamefont{Liu}},
  \bibinfo{author}{\bibfnamefont{S.~J.} \bibnamefont{Chung}},
  \bibinfo{author}{\bibfnamefont{K.~J.} \bibnamefont{Goldammer}},
  \bibnamefont{and} \bibinfo{author}{\bibfnamefont{M.~B.}
  \bibnamefont{Santos}}, \bibinfo{journal}{Physica E}
  \textbf{\bibinfo{volume}{6}}, \bibinfo{pages}{293} (\bibinfo{year}{2000}).

\bibitem{remark1}
 \bibinfo{note}{Due to strong $\delta$-doping with Be resp. Si the
  density of states becomes asymmetric and the critical filling factor is
  shifted from the ideal $\nu_c=1.5$ to lower resp. higher filling factors},
 \bibinfo{author}{\bibfnamefont{R.~J.} \bibnamefont{Haug}},
  \bibinfo{author}{\bibfnamefont{R.~R.}~\bibnamefont{Gerhardts}},
  \bibinfo{author}{\bibfnamefont{K.}~\bibnamefont{v.~Klitzing}},
  \bibnamefont{and} \bibinfo{author}{\bibfnamefont{K.}~\bibnamefont{Ploog}},
  \bibinfo{journal}{Phys. Rev. Lett.}
  \textbf{\bibinfo{volume}{59}}(\bibinfo{number}{12}), \bibinfo{pages}{1349}
  (\bibinfo{year}{1987}{\natexlab{a}}).

\bibitem[{\citenamefont{Ploog}(1987)}]{ploog87}
\bibinfo{author}{\bibfnamefont{K.}~\bibnamefont{Ploog}}, \bibinfo{journal}{J.
  Cryst. Growth.} \textbf{\bibinfo{volume}{81}}, \bibinfo{pages}{304}
  (\bibinfo{year}{1987}).

\bibitem[{\citenamefont{Girvin and MacDonald}(1997)}]{girvin97}
\bibinfo{author}{\bibfnamefont{S.~M.} \bibnamefont{Girvin}} \bibnamefont{and}
  \bibinfo{author}{\bibfnamefont{A.~H.} \bibnamefont{MacDonald}}, in
  \emph{\bibinfo{booktitle}{Perspectives in Quantum {Hall} Effects}}, edited by
  \bibinfo{editor}{\bibfnamefont{S.} \bibnamefont{Das Sarma}} \bibnamefont{and}
  \bibinfo{editor}{\bibfnamefont{A.}~\bibnamefont{Pinczuk}}
  (\bibinfo{publisher}{Wiley}, \bibinfo{address}{New York},
  \bibinfo{year}{1997}).


\end{thebibliography}
\end{document}